\documentclass[aps,pra,floatfix,nofootinbib]{revtex4}
\usepackage{graphicx}
\usepackage[english]{babel}
\usepackage{amsmath}
\usepackage{amssymb}
\usepackage[usenames,dvipsnames]{color}
\usepackage{color}
\definecolor{Blue}{rgb}{0.3,0.3,0.9}
\definecolor{orange}{rgb}{1,0.5,0}
\usepackage{subfigure}
\usepackage{bm}
\usepackage{titlesec}

\definecolor{refkey}{rgb}{0.6,0,1}
\definecolor{labelkey}{rgb}{0.6,0,1}

\newcommand{\csch}{{\rm \, csch}}
\newcommand{\sech}{{\rm \, sech}}
\newcommand{\gd}{{\rm gd}}

\newcommand{\mfmadd}[1]{{#1}}

\newcommand{\rjadd}[1]{{ #1}}

\begin{document}
\title{A diagrammatic expansion of the Casimir energy in multiple reflections: theory and applications}
\author{Mohammad F. Maghrebi}
\affiliation{Center for Theoretical Physics, and Department of Physics, Massachusetts Institute of Technology, Cambridge, MA 02139, USA}
\begin{abstract}
We {develop} a diagrammatic representation of the Casimir energy of a multibody configuration. The diagrams represent multiple reflections between the objects and can be organized by a few simple rules. The lowest-order diagrams (or {\it reflections}) give the main contribution to the Casimir interaction which proves
 the usefulness of this expansion. Among some applications of this, we find analytical formulae describing the interaction between {``edges'', {\it i.e.\/} semi-infinite plates,} where we also give a first example of {\it blocking} in the context of the Casimir energy. We also find the interaction of edges with \mfmadd{a needle} and describe analytically a recent model of the {\it repulsion} due to the Casimir interaction.
\end{abstract}
\maketitle

\section{Introduction}

In 1948, Casimir predicted an attractive force between metal plates arising from quantum fluctuations \cite{Casimir48-2}. The advent of experimental measurements of Casimir forces has stimulated a large interest in this field \cite{Mohideen98,Harris00,Bressi02}. \mfmadd{There have been extensive studies of the Casimir force both analytically \cite{Milton01, Kenneth06, Emig07, Bordag09} and numerically \cite{Gies06, Reid09, Pasquali09}}. \mfmadd{Specifically,} a multipole scattering method has been developed and used to compute this force between multiple objects of {various} shapes and electromagnetic properties \cite{Emig07,Rahi09}.
This formalism allows one to compute the Casimir interaction in a multiple scattering scheme based on the scattering matrix of each object. The conceptual foundations of this approach can be traced back to earlier multiple scattering formalisms \cite{Balian77, Balian78}.

\mfmadd{Within this formalism, we should compute the logarithm of the determinant (the $\ln\det$ formula) of a certain matrix in order to compute the energy \cite{Kenneth06, Emig07}.}
This can be expanded in a series of {\it multiple reflections}. This is a particularly useful expansion in the Casimir-Polder limit for small-sized objects.
In this limit, the first reflection gives the leading order contribution while higher reflections (and partial waves) are suppressed by higher powers of the objects' length scale divided by the separation distance. It is not immediately clear that this would be a useful, or even sensible, expansion for a more general configuration far from this limit.
However, a recent work \cite{Maghrebi10} on the Casimir force between non-compact objects (cones, wedges and plates) necessitated the use of such expansion: in this application the partial waves are labeled by a {\it continuous} set of quantum numbers and thus it is not convenient to use the $\ln\det$ formula\mfmadd{, which involves the determinant of a continuously labeled matrix,} in its general form. The multiple reflections captures the $\ln\det$ formula as a sum over the \mfmadd{trace of certain operators which is well-defined for continuous indices.}
Remarkably, this expansion enjoys a rapid convergence in the number of multiple reflections regardless of any geometrical limit.
This observation has motivated this work in which we elaborate in detail on the derivation and applications of multiple-reflection expansion for multibody configurations.

This paper is organized as follows. In Sec.~\ref{Sec. Multiple reflections: derivation and discussion}, we derive a diagrammatic expansion of the multibody Casimir interaction which is shown to be organized by a few simple rules. \mfmadd{The convergence behavior of this expansion is discussed in Sec. \ref{Sec. Convergence}. }Finally in Sec.~\ref{Sec. Applications}, we propose some applications. As an example, we consider the interaction between edges and find very accurate analytical formulae in Sec.~\ref{Sec. Interaction of edges}.
We study the interaction between edges \mfmadd{and a needle} in Sec.~\ref{Sec. Interaction of edges and dipoles} where we {again} find explicit analytical formulae. This is used to describe a repulsive Casimir interaction for a specific configuration of the edges and \mfmadd{the needle}.

\section{Multiple reflections: derivation and discussion}\label{Sec. Multiple reflections: derivation and discussion}
\begin{figure}[h]
 \centering
 \def\svgwidth{8cm}
 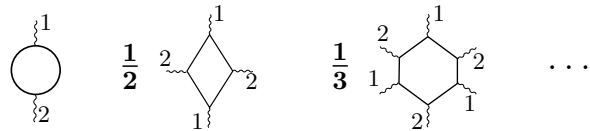
 \caption{\label{Fig-TwoBodyDiagExp}Diagrammatic expansion of the Casimir energy for two objects. The vertices label the objects and the links represent the translation matrices between them. A diagram with $n$ insertions of 1 and 2 comes with a factor of $1/n$. The dots represent higher order diagrams.}
\end{figure}

The Casimir interaction energy between two objects can be computed by using functional integral methods. In this approach, the interaction can be described by {charge and current} multipoles which exist on each object, and are later integrated out to obtain an expression for the energy only in terms of the scattering properties of the objects. For example, for two objects \cite{Kenneth06, Emig07}
\begin{equation}\label{Eq. Trlog two body energy}
  \mathcal{E} = \frac{\hbar c}{2 \pi}  \int_{0}^{\infty} d\kappa\, {\rm tr}\ln\left(\mathcal I- T_1 \, \mathcal{U}_{12}T_2\, \mathcal{U}_{21}\right),
\end{equation}
\mfmadd{where we replaced the logarithm of the determinant ($\ln\det$) by the trace of the logarithm ($\rm tr \ln$) for the sake of convenience in the following discussion.}
Here $T_1$ and $T_2$ are the scattering matrices or the $T$-matrices which encode all properties of the objects including their shape (and their electromagnetic properties in the case of electrodynamics). The matrices $\mathcal U_{12}$ and $\mathcal U_{21}$ are the translation matrices which capture the appropriate translations and rotations between the scattering bases for each object. Equation~(\ref{Eq. Trlog two body energy}) in this form is exact and can be used to compute the Casimir forces between any two objects. However, as outlined in the Introduction, there are cases where we have to expand the $\rm trln$ formula as a power series in $\mathcal N =T_1 \,\mathcal{U}_{12}T_2 \, \mathcal{U}_{21}$
\begin{equation}\label{Eq. Trlog two body expansion}
  \mathcal E = -\frac{\hbar c }{2 \pi} \int_{0}^{\infty} d \kappa\left(\rm tr \mathcal N+\frac{1}{2}\rm tr \mathcal N^2+\frac{1}{3}\rm tr \mathcal N^3+\cdots\right).
\end{equation}
This finds a simple interpretation by noting that each term includes one more reflection (back and forth) between the two objects. Diagrammatically this can be represented as in {Fig.~\ref{Fig-TwoBodyDiagExp}}. The number in front of each diagram is the coefficient of this term in expanding the logarithm.
This expansion has been used to obtain analytical formulae for the interaction of wedges and cones with plates \cite{Maghrebi10}. Indeed the rapid convergence required only the computation of the lowest few reflections. We study the convergence in a variety of cases in the next sections, but in the following discussion we focus on the derivation of multiple-reflection expansion in general.
In principle, for three or more objects, we can find a ${\rm tr} \ln$ expression which involves the scattering matrices of all objects but in a somewhat complicated form. An expression for the Casimir energy for three objects, for example, is {given in Eq.~(III.27)} of Ref.~\cite{Emig07}.  For more objects, the expression for the energy will be increasingly more complicated.

\begin{figure}[h]
 \centering
\def\svgwidth{13cm}
 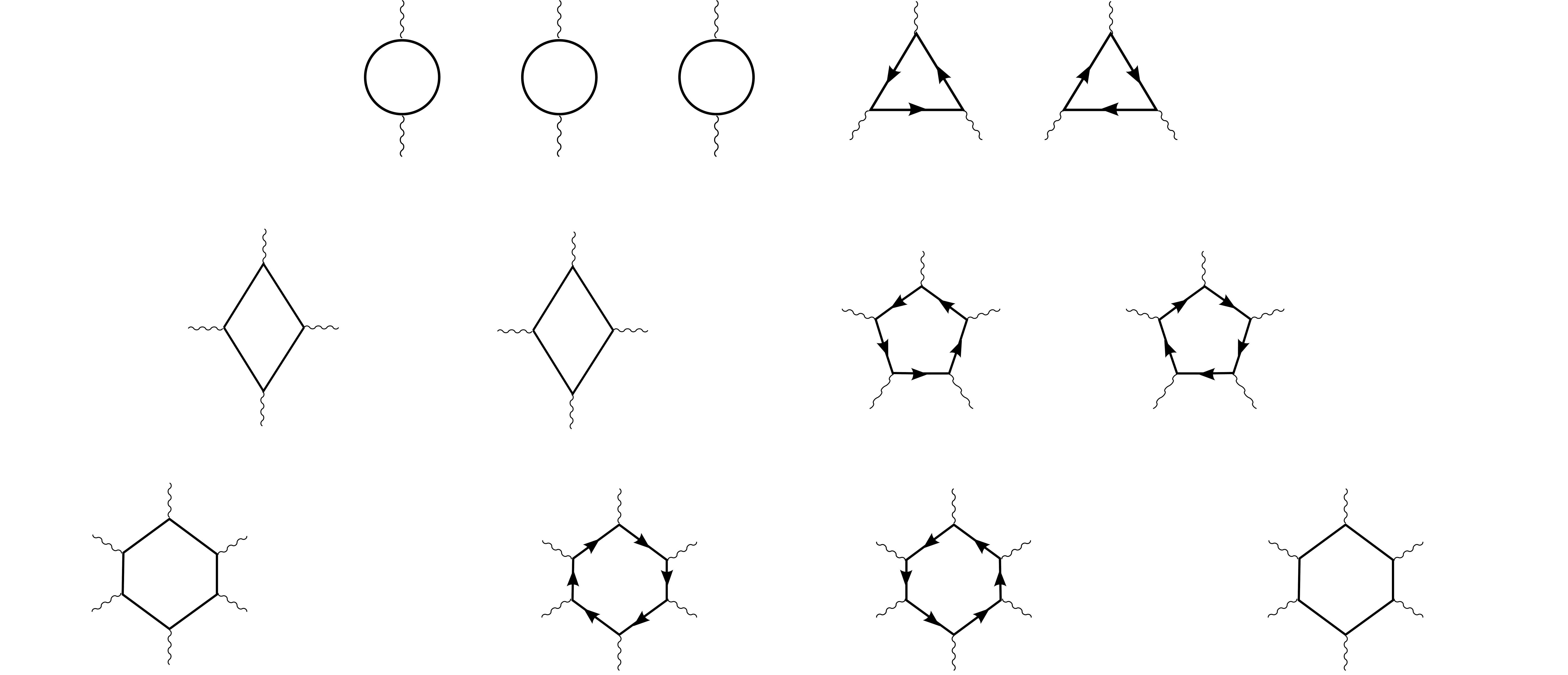
 \caption{Diagrammatic expansions of the Casimir energy for three objects. The dots represent either a permutation of the objects in the same diagrams or higher-order reflections. \mfmadd{Each diagram is accompanied with a numerical factor (1 if not stated explicitly). The diagrams are listed in the order of the number of reflections. The diagrams $[321]$ and [231], for example, are not identical, so they are both included while $[3121]$ \rjadd{is identical to $[2131]$ (after a cyclic transformation) and is therefore omitted.  Furthermore the loop in $[2131]$ can be run either way without changing the diagram (up to cyclic permutations),}  thus the corresponding diagram does not have an arrow.}} \label{Fig-ThreeBodyDiagExp}
\end{figure}

Nevertheless, the ${\rm tr} \ln$ formula can be expanded and organized in multiple reflections. For three objects, for example,
the first few diagrams
are listed in {Fig~\ref{Fig-ThreeBodyDiagExp}}.
\\
\mfmadd{One can see that the diagrammatic expansion is organized by a few simple rules. First we state the rules for the Casimir interaction of an arbitrary number of objects. Later in this section we derive these rules systematically.}

\mfmadd{Let us suppose that there are $M$ objects which are separated from each other in the vacuum. Each diagram forms a closed loop which consists of a certain number of insertions of the objects, and these are drawn by the wavy lines. The links between the objects denote the translation matrix. We are interested in the {\it interaction} of multiple objects, so we ignore the diagrams with a single insertion of the $T$-matrix. To define our set of rules, we introduce a more abstract notation. Consider a diagram with $N$ insertions of the objects: we designate the objects by $i_1, i_2, \dots, i_N$ in the order that they appear in the diagram and represent the latter by $[i_N \dots i_2 i_1]$. Since the diagrams are closed loops, they have cyclic symmetry. So, for example, $[i_N \dots i_2 i_1]$ and $[i_{N-1} \dots i_1 i_N]$ are identical. Note that the order that the objects are listed is important. In general $[i_N i_{N-1} \dots i_1]$ is different from $[i_1 i_2 \dots i_N]$.} The multiple-reflection expansion is organized by the following rules:

1. Only {\it different} objects are connected by a link; an object index is not succeeded by the same index. \mfmadd{So, for a diagram defined by $[i_N \dots i_2 i_1]$, $i_k \ne i_{k+1}$ where $N+1$ is identified with 1.}

2. \rjadd{ \mfmadd{A diagram defined by $[i_N \dots i_2 i_1]$ is symmetric if it is identical to $[i_1 i_{2} \dots i_{N}]$ up to a cyclic transformation.}  The diagrams which are symmetric in both directions (of the loops) do not have an arrow and appear only once while those which are not, appear in pairs of opposite directions.}

3. \mfmadd{Let us define a subdiagram as a subset of the diagram which is interconnected.} A diagram which is made from $n$ copies of the same subdiagram comes with a combinatoric factor of $1/n$. \mfmadd{For example, a diagram represented by $[i_2 i_1i_2 i_1 \dots i_2 i_1]$ with $n$ copies of $i_2 i_1$ should be multiplied by $1/n$.}\\
\\
\mfmadd{In short, a diagram defined by $[i_N \dots i_2 i_1]$ corresponds to the following term in the expansion of the energy
\begin{equation}
  \mathcal E_{[i_N \dots i_2 i_1]}=-\frac{\hbar c }{2 \pi}\, S_{[i_N \dots i_2 i_1]}\int d\kappa  \, {\rm tr}\left( \mathcal U_{i_1 i_N}T_{i_N} \mathcal U_{i_{N}i_{N-1}}\dots T_{i_2}\mathcal U_{i_2 i_1}T_{i_1}  \right),
\end{equation}
where the $T$ and $\mathcal U$ represent the scattering and translation matrices respectively. The constant $S$ is the symmetry factor of the corresponding diagram and is determined by Rule 3.
}
\\
\\
These can be proved in general.
We start from the partition function whose logarithm is proportional to the energy\footnote{\mfmadd{We set the temperature to zero for convenience. The generalization to finite temperature is straightforward.}}.
\mfmadd{Using path-integral techniques, it is possible to express the partition function as a functional integral over {charge-current} multipoles, $Q_\alpha$. This has been done in great detail in Ref.~\cite{Emig07}.} The translational invariance in time allows one to express the partition function as a product of all frequencies, $\kappa$,\footnote{The exponent, or the {\it action}, is always real because the scattering matrix $T(i\kappa)$ and the translation matrix $\mathcal U(i\kappa)$ are Hermitian for imaginary frequency.}
\begin{equation}\label{Eq. Partition function}
  \mathcal Z= \prod_{\kappa>0}\prod_{\alpha=1}^{N} \int_{0}^{\infty} [\mathcal DQ_\alpha \mathcal DQ_\alpha^*]\exp\left\{-\frac{\kappa}{2}\frac{t_0}{\hbar} \sum_\alpha Q_\alpha^*[T_\alpha]^{-1}Q_\alpha +\frac{\kappa}{2}\frac{t_0}{\hbar} \sum_{\alpha\ne \beta} Q_\alpha^* \, \mathcal U_{\alpha \beta}Q_\beta\right\}.
\end{equation}
\mfmadd{where it is evaluated in a time interval $t_0$. Note that $T_\alpha$ is the scattering matrix of the object $\alpha$ and $\mathcal U_{\alpha\beta}$ is the translation between the objects $\alpha$ and $\beta$.}
When integrated over the multipoles, the logarithm of the last equation gives the ${\rm tr}\ln$ formula (up to a normalization factor). Instead of performing the functional integral, we will proceed as follows. The action (or the exponent of the integrand of the functional integral) defines a natural set of Feynman rules ---by which we can organize the diagrammatic structure of the Casimir energy.

\mfmadd{Heretofore, we consider the $T$-matrices to represent the vertices in the diagrams as they naturally describe the interaction (of the electromagnetic waves) with the objects. Similarly, the translation matrices are represented by the links or the propagators which connect separate objects. For the sake of derivation of the diagrammatic rules, we consider a different set of conventions in the following discussion. In the rest of this paper, we use the same conventions as we defined earlier.}
\begin{figure}[ht]
\centering
\def\svgwidth{8cm}
\subfigure[{\label{Fig-FeynmanRules} {\it Feynman} {\it propagator} and {\it vertex}.}]{
 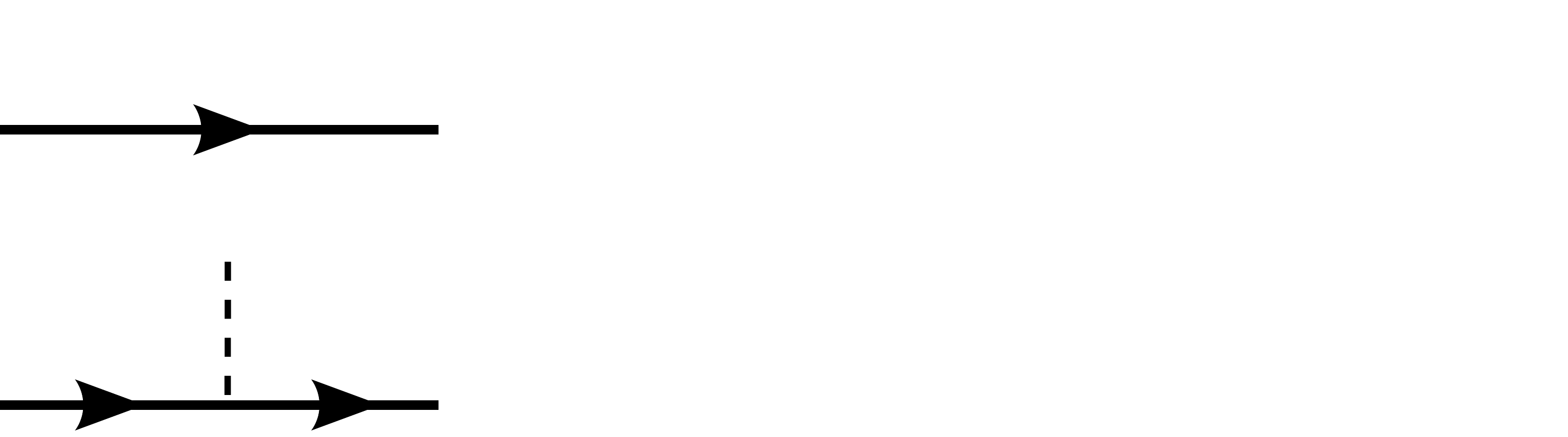
}\qquad
\def\svgwidth{4cm}
\subfigure[\label{Fig-AFeynmanDiagram} {{A {\it Feynman} diagram.}}]{
 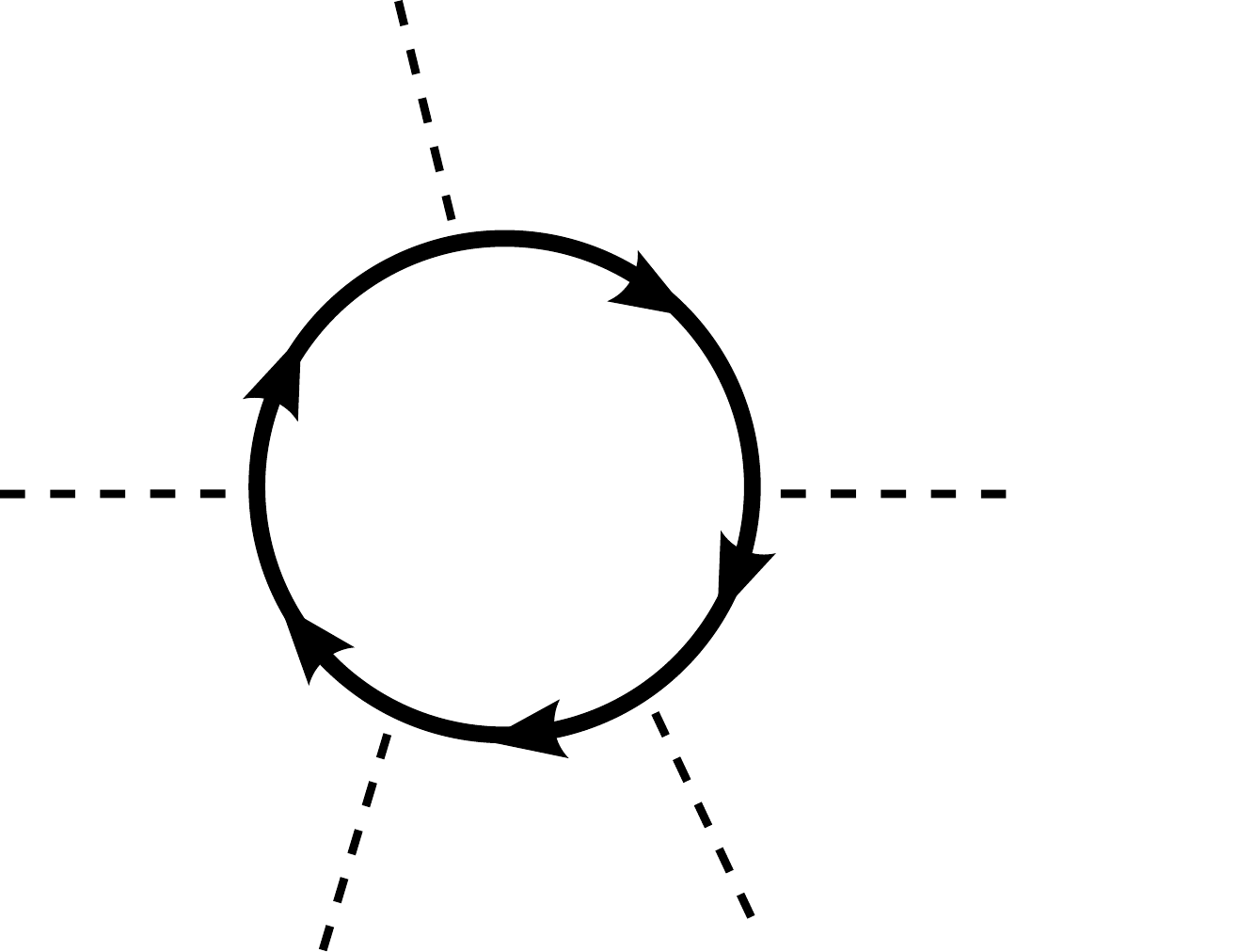
\label{Fig-AFeynmanDiagram}
}
\caption{{Feynman rules for the Casimir energy. The diagram on the right corresponds to the term $-\frac{\hbar c}{2 \pi}\int d\kappa \,{\rm tr} (  T_3\,\mathcal U_{31}T_1\,\mathcal U_{13}T_3\,\mathcal U_{32}T_2\,\mathcal U_{21}T_1 \,\mathcal U_{13})$ in the expansion of the $\rm trln$ formula.}}
\end{figure}\\
Note that the first term in Eq. (\ref{Eq. Partition function}) is diagonal in the object indices.
\mfmadd{So, from a mathematical point of view,} it is natural to consider it as a {\it Feynman propagator}.
The propagator is the inverse kernel, hence we must attribute a factor of $T_\alpha$ to any propagating line\footnote{Factors of $\frac{\kappa T}{2\hbar}$ can be absorbed in the definition of the multipoles $Q_\alpha$ and do not play a role here.}. Also, the multipoles are complex, so we must associate a direction ---or an arrow--- to the propagator. The second term in the exponent only connects the multipoles on separate objects (it is off-diagonal in the object indices) and so can be more appropriately considered as an {\it interaction vertex}. These rules are depicted in {Fig.~\ref{Fig-FeynmanRules}}. A typical diagram made from \mfmadd{the latter }rules is given in {Fig.~\ref{Fig-AFeynmanDiagram}}.

\mfmadd{These diagrams are dual to the diagrams in Figs.~\ref{Fig-TwoBodyDiagExp} and \ref{Fig-ThreeBodyDiagExp}, {\it i.e.,} the objects are represented by lines and the translation matrices by the dashed external lines due to the different choice of conventions.} The three rules which were stated earlier follow straightforwardly. Rule 1 follows trivially because the translation matrix is off-diagonal. Rule 2 holds because there is an arrow on the lines. Therefore the two triangular diagrams in Fig.~\ref{Fig-ThreeBodyDiagExp}, for example, are not equivalent\footnote{By hermiticity of the $T$-matrices and the translation matrices, one can easily see that the two triangular diagrams are related by complex conjugation. The energy is real because it involves the sum of these diagrams.}. The same arrow can be drawn on the symmetric diagrams, however, the two directions of the arrow generate the same diagram and must be counted only once. For these diagrams, the arrow is irrelevant and can be dropped. Finally Rule 3 follows because the cyclic symmetry of diagrams requires a symmetry factor of $1/n$.

\rjadd{After discussing the convergence of the expansion in the following section,} we will give some applications of the multiple-reflection expansion. To show the strength of this method, we consider the interaction of edges among themselves and with dipoles. The analytical results we will find should be of both theoretical and experimental interest.
\mfmadd{
\section{Convergence}\label{Sec. Convergence}
In the previous section, we showed that, at least formally, the Casimir energy can be expanded in a series of multiple reflections, or {\it diagrams}, which can be organized by a few simple rules. In this section, \rjadd{although lacking a general proof,}  we argue that under very generic circumstances the lowest orders of reflections dominate the Casimir energy.

Casimir computed the force between two perfectly reflecting parallel plates \cite{Casimir48-2}. A simple extension to $D$ spatial dimensions generalizes the Casimir force to
\begin{equation}\label{Eq. Parallel Plates Force}
  F=-\frac{a_D \, \hbar c A}{d^{D+1}} \zeta[D+1] .
\end{equation}
In this equation, $A$ is the area of the plates, $d$ is their separation distance, $\zeta(D+1)$ is the zeta function and $a_D$ is a constant which depends on the dimensionality. It is instructive to compute the force order by order in multiple reflections. The contributing diagrams are given in Fig.~\ref{Fig-TwoBodyDiagExp}. The $T$-matrix for the (perfectly reflecting) plates is $T_{M/E}(i \kappa)=\mp 1$ where $\kappa$ is the imaginary frequency, and the $T$-matrix of the magnetic (M) and the electric (E) polarizations correspond to the upper and lower signs respectively. One can then easily see that a diagram with $n$ insertions of the plate 1 and $n$ insertions of the plate 2 contributes to the energy by
\begin{equation}\label{Eq. Parallel Plates Force in multiple reflections}
  F_n=-\frac{a_D \, \hbar c A}{d^{D+1}}\frac{1}{n^{D+1}}.
\end{equation}
\rjadd{This result was noted earlier in the application of the optical approximation to  parallel plate geometry, where the ``reflections'' are literally the specular reflections of ray optics \cite{Scardicchio:2004fy}}.
So the contribution of the higher reflections to the force falls off in a power-law fashion. In three dimensions, the latter converges as $1/n^4$. If the plates are dielectric as opposed to perfect reflectors, the multiple-reflection expansion of the force converges even more rapidly. Note that the $T$-matrix of a plate remains diagonal \mfmadd{(in the planar wave basis)} at finite conductivity. The absolute value of the eigenvalues of the $T$-matrix can be shown to be less than (or equal to) unity for any dielectric response, {\it i.e.,} $|T_{M/E}(i\kappa)|\le 1$ at imaginary frequency. The latter can be used to find a bound on the contribution of the $n$-th order diagram as expanded in Fig.~\ref{Fig-TwoBodyDiagExp},
\begin{equation}
  F_n \le \frac{1}{n^{D+1}} F_1.
\end{equation}

Casimir's computation of perfectly reflecting plates can also be altered by changing the geometry while keeping the perfect reflectivity intact. Let us consider a wedge opposite a plate in three dimensions, which was extensively studied in Ref.~\cite{Maghrebi10}. The expansion in multiple reflections is found to be governed approximately by a power law $1/ n^{4+\delta}$ for a positive $\delta$ which is a function of the opening angle and the orientation of the wedge.
So the convergence is more rapid than parallel plates. The perfectly reflecting plate-plate and wedge-plate configurations are both scale invariant, {\it i.e.}, their only length scale is the separation distance between the two objects. Therefore, the convergence in multiple reflection is not controlled by any dimensionless factor and is purely numerical. Now we consider a geometry which provides an internal length scale. The interaction of a sphere of radius $R$ with a plate at a distance $d$ in three dimensions can be studied in various regimes. If the separation distance is small ($d\ll R)$, {\it proximity force approximation} (PFA) can be used to compute the force \cite{Derjaguin56}. This approximation is based on treating the objects locally as parallel plates and integrating over the surfaces facing each other by using Eq.~(\ref{Eq. Parallel Plates Force}). In the close proximity limit, PFA becomes exact. So by analogy with parallel plates, the convergence is governed by $1/n^4$ series in the limit of small sphere-plate separation. In the opposite limit where the sphere is small ($d\gg R$), the energy is dominated by the dipole interaction \cite{Casimir48-1}. In fact, the lowest reflection in the leading order of partial waves gives a force which is proportional to $R^3/ d^3$. The contribution of the higher reflections is exponentially suppressed: the $n$-th order diagram as expanded in Fig.~\ref{Fig-TwoBodyDiagExp} becomes proportional to $(R^3/d^3)^n$ in the leading order of partial waves.

In summary, the expansion in multiple reflections appears to converge rapidly: in close proximity (the absence of a small parameter), the convergence is governed by a power law and is purely numerical whereas in the opposite limit, \rjadd{the convergence is exponential} and is controlled by the ratio of the object's length scale divided by the separation distance. Intuitively, the convergence in reflections can be understood as follows. More reflections typically travel over a larger optical path. Note that in the $n$-th reflection, the length of a typical path traveled by the waves between two parallel plates is $\ell_{\rm path}\sim 2 n d$. The force in the $n$-th reflection is, in fact, proportional to the $1/\ell_{\rm path}^4$ (see Eq.~(\ref{Eq. Parallel Plates Force in multiple reflections})) \cite{Scardicchio:2004fy}.
In general, the optical path increases for higher reflections. This intuition serves as our guiding principle in deciding which reflections contribute the most. The Casimir interaction of a multibody configuration can be organized by a similar expansion of multiple reflections as discussed in Sec.~\ref{Sec. Multiple reflections: derivation and discussion}. In various examples that we study in the following section, we consider the reflections which correspond to the shortest optical paths and compute the Casimir interaction. We confirm the latter criterion by computing the higher orders in reflections and show that their contribution is indeed much smaller than the lowest reflections.
}

\section{Applications}\label{Sec. Applications}
\subsection{Interaction of edges}\label{Sec. Interaction of edges}
In this section, we consider the interaction between the edges {of half plates}.
In Sec.~\ref{Sec. Two half-plates}, we find an analytical expression for the interaction of two edges which can be studied in various limits and configurations. In Sec.~\ref{Sec. Three half-plates}, we consider three half-plates and point to some of its interesting physical behaviors, including the non-monotonic dependence of the force on the separation distance. In Sec.~\ref{Sec. Blocking}, we study a first example of {\it blocking} in the context of the Casimir energy.

\subsubsection{Two half-plates}\label{Sec. Two half-plates}
\begin{figure}[h]
  \centering
  \def\svgwidth{5.5cm}
  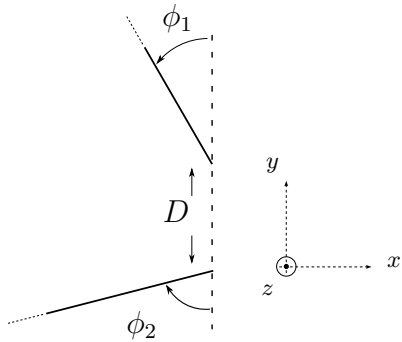
\caption{\label{Fig-TwoHalfPlates} {The configuration of two half-plates.}}
\end{figure}

This study of the interaction between two semi-infinite plates reproduces same results reported in Ref.~\cite{TBP1}. \mfmadd{We assume that the two half-plates are parallel along their edge and} define the angle that they make with the edge-to-edge axis to be $\phi_1$ and $\phi_2$ (see Fig.~\ref{Fig-TwoHalfPlates}). \mfmadd{The half-plates are separated by a distance $D$, and their extent along the axis $z$ (perpendicular to  Fig.~\ref{Fig-TwoHalfPlates}) is $L\gg D$}. The scattering matrix of a half-plate depends only trivially on the axial wavevector\mfmadd{, the component of the wavevector parallel to the edges which we denote by $k_z$}. \mfmadd{It depends on the wavevector of the incoming plane wave $\vec k$ as well as the scattered plane wave $\vec k'$. Both the incoming and scattered waves are defined at imaginary frequency $\kappa$. The $T$-matrix is given explicitly in terms of the wavevectors by} \cite{Maghrebi10}
\begin{equation}\label{Eq. half plate T-matrix}
  T^{D/N}_{\kappa \vec k',\kappa \vec k}=\frac{1}{2} \left(-\sec\left(\frac{{a'}^*-a}{2}\right)\mp \sec\left(\frac{{a'}^*+a}{2}\right)\right) \mfmadd{2\pi\delta({k_z-k_{z'}})},
\end{equation}
where $a$ is the angle that the incident plane wave makes with the half-plate while $a'$ is the angle of the scattered wave, \mfmadd{{\it i.e.,} $a= \sin^{-1}\left(\frac{k_x}{i\kappa}\right)$ and $a= \sin^{-1}\left(\frac{k'_x}{i\kappa}\right)$}.
The upper and lower signs correspond to Dirichlet (D) and Neumman (N) boundary conditions respectively.

The wavenumber is imaginary so the angles are in general complex.
Let us choose the line which connects the two edges to be the vertical axis. We then define $a=i\alpha-\phi_j$ and $a'=i\alpha'-\phi_j$ where $j=1, 2$ labels the first or the second half-plate respectively. Note that $i\alpha$ and $i\alpha'$ are the (imaginary) angles that the waves make with the vertical axis.

The $T$-matrix in Eq.~(\ref{Eq. half plate T-matrix}) was constructed in Ref.~\cite{Maghrebi10} to find the interaction of a half-plate and an infinite perfectly conducting plate. By incorporating the $T$-matrices into the translation matrices \cite{Rahi09},
we can find the Casimir energy through Eq.~(\ref{Eq. Trlog two body energy}). Since the indices of the $T$-matrices are continuous, it is more convenient to expand the trln formula. We compute the Casimir energy in the first reflection ({\it i.e.}, the first diagram in Fig.~\ref{Fig-TwoBodyDiagExp}).
An analytical expression can be obtained for Dirichlet, Neumann or electromagnetic boundary conditions
\begin{align}
  \mathcal E^{D/N}= -\frac{\hbar c L }{128\pi^3D^2}\Big(
  &\pm\frac{\pi}{2} \left(\cos^2\phi_1\left(\pi-\pi \cos\phi_1+\gd(i\phi_1)+\gd(-i\phi_1) \right)+\phi_1\rightarrow \phi_2\right) \nonumber \\
  &+\frac{8}{3} -4 \cos \phi_1 \cos\phi_2+4\left(\phi_1 \csc^2\phi_1+\phi_2 \csc^2\phi_2\right) \csc(\phi_1+\phi_2)\Big).
\end{align}
Here $\rm gd$ is the Gudermannian function. Because of the translational symmetry in one direction, the electromagnetic Casimir energy is simply the sum of Dirichlet and Neumann results. The range of validity of this expression is all $|\phi_1|, |\phi_2| \le \pi/2$ but it can be analytically continued to the regime where one of the angles or both exceed $\pi/2$.

The first reflection computation gives a very good approximation to the exact result. This has been studied in some detail in Ref.~\cite{TBP1} for various limits and configurations.

\subsubsection{Three half-plates}\label{Sec. Three half-plates}
In this section, we study a three-body interaction; specifically a configuration of three half-plates. We
\begin{figure}[h]
  \centering
  \def\svgwidth{5cm}
  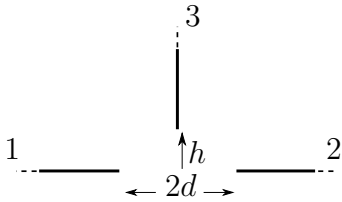
\caption{\label{Fig-ThreeHalfPlates} The configuration of three half-plates.}
\end{figure}
will demonstrate the convergence in multiple reflections and show that to the lowest order in reflections, and for a wide range of configurations, the sum of the leading diagrams dominates the Casimir energy.

Let us take two half-plates aligned but extended in opposite directions while keeping the third half-plate perpendicular to them along the axis midway between
the two ({see Fig.~\ref{Fig-ThreeHalfPlates}}). We find the force which is exerted on the vertical half-plate. Obviously the force {goes to} zero at large separation  distance (large positive $h$). It also vanishes when the vertical half-plate goes far below the other two. The latter is the limit that an infinite-plate is inserted between the two half-plates in which case the lateral force on the infinite plate vanishes. Given the limits, the force cannot be {a monotonic function of $h$}.

\begin{figure}[h]
  \centering
  \def\svgwidth{10cm}
  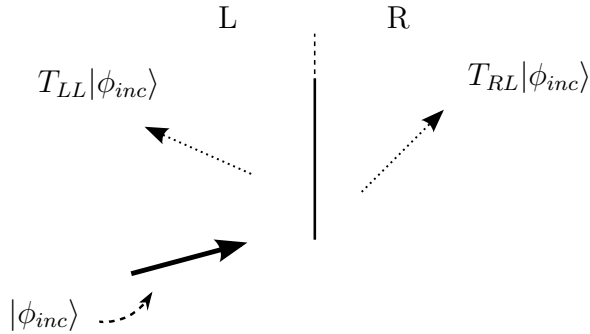
\caption{\label{Fig-TmatChannels} The left-left and right-left channels of the $T$-matrix.}
\end{figure}

In order to {compute} the force, we {first} construct the $T$-matrices. However, there are two different channels of the $T$-matrix which come into play. When an incoming wave impinges on the half-plate from one side, the scattered wave propagates in two directions, {\it i.e.}, back to the same side and forward to the opposite side. We call the two channels, the LL and RL channels for left-to-left and left-to-right scattering\footnote{By symmetry, the RR and RL channels are identical to LL and LR channels, respectively.} (see {Fig.~\ref{Fig-TmatChannels}}).
The LL channel is similar to what we considered earlier: the incident wave angle is $a=i\alpha-(\pi/2-\epsilon)$ while the scattered wave angle is $a'=i\alpha'-(\pi/2-\epsilon)$. \mfmadd{This is a special case of the setup in Fig.~\ref{Fig-TwoHalfPlates} where the axis $y$ makes a right angle with the half-plate and} $\epsilon$ is a small positive number which is essential for the regularity of the solutions. In the RL channel the scattered angle becomes $a'=-i\alpha'+(\pi/2-\epsilon)$. Putting these in Eq.~(\ref{Eq. half plate T-matrix}), we find
\begin{align}
  & T^{D/N}_{LL\, \alpha', \alpha}=\frac{1}{2} \left( \mp \sech\left(\frac{\alpha+\alpha'}{2}\right)-i \csch\left(\frac{\alpha-\alpha'}{2}+i \epsilon\right)\right) \mfmadd{2\pi\delta({k_z-k_{z'}})},\\ \label{Eq. LL/RL symmetry}
  & T^{D/N}_{RL\, \alpha', \alpha}=\pm T^{D/N}_{LL\, \alpha', \alpha}\,.
\end{align}
Interestingly the two channels are related simply by a sign. It can be shown that this holds for any planar geometries, {\it i.e.}, for any geometry which is part of a flat screen. This, among other applications, is discussed in Ref.~\cite{TBP2}.

\begin{figure}[h]
  \centering
  \def\svgwidth{10cm}
  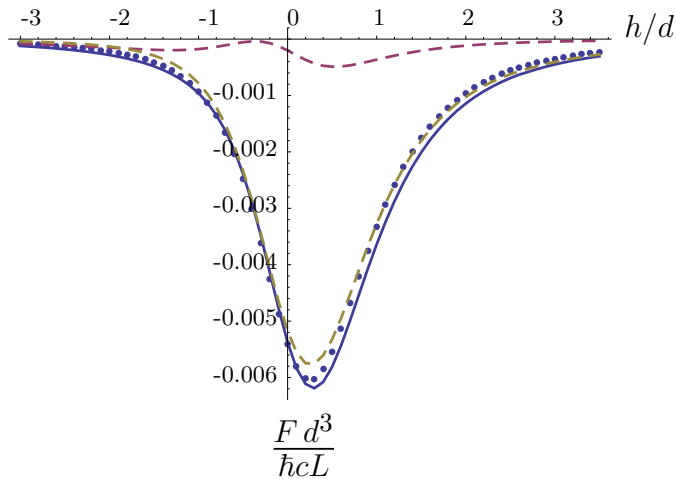
\caption{\label{Plot-ForceOnVerticalHalfPlate} The force on the vertical half-plate as a function of the separation distance.  \mfmadd{The solid curve represents the electromagnetic force computed by summing the two-body diagrams ($[21]$ and $[31]$) only. The dots include corrections from higher reflections. The dashed curves represent the force corresponding to the Dirichlet and Neumann boundary conditions.}}
\end{figure}

\begin{figure}[h]
  \centering
  \def\svgwidth{10cm}
  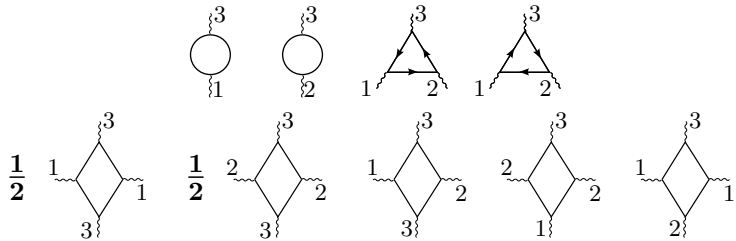
\caption{\label{Fig-ForceThreePlate} The lowest-order diagrams in reflections contributing to the vertical force. \mfmadd{The diagrams correspond to the shortest optical paths between the objects. The first two diagrams \mfmadd{($\ell_{\rm path}\sim 2d$)} dominate the Casimir energy. The rest of these diagrams \mfmadd{($4 d\lesssim \ell_{\rm path}\lesssim 6d$)} give small corrections to the leading order contribution. }}
\end{figure}

The force on the vertical half-plate is plotted in {Fig.~\ref{Plot-ForceOnVerticalHalfPlate}}.
The {solid (blue) curve} gives the electromagnetic force which is computed in the lowest reflection, {\it i.e.}, as the sum of the two-body forces between the vertical half-plate and either of the two horizontal half-plates (the first two diagrams in Fig.~\ref{Fig-ForceThreePlate}). The {dotted curve} shows the force which is summed up to the first few orders shown in {Fig.~\ref{Fig-ForceThreePlate}} (including the two-body terms).
The two dashed curves give the energy in the first reflection for Dirichlet and Neumann boundary conditions. The  Neumann result is smaller than {Dirichlet} by nearly one order of magnitude.

{Interestingly, we see that the maximum of the force happens before the vertical half-plate reaches the axis along the horizontal ones, {\it i.e.}, the force is not monotonic even when the third half-plate \mfmadd{is above the horizontal half-plates}.}

It is clear that the convergence in multiple-reflection expansion is remarkably good for a wide range of {the parameter $h$}. Note that we have not taken advantage of any geometrical approximation such as the Casimir-Polder limit; nor have we ignored the nonadditivity of the Casimir force by resorting to an approximation like PFA. Nevertheless, we can find the force {to quite good accuracy}.

\subsubsection{Blocking} \label{Sec. Blocking}
In the previous section, we computed the force on the vertical semi-infinite plate. Here, we consider a similar setup but focus on a different physical quantity, namely
how the presence of the third object influences the interaction between the other two. Suppose that the horizontal {distances of the half-plates 1 and 2 to the vertical half-plate are} $d_1$ and $d_2$ respectively. ({We set these equal to $d$ later on}.) The horizontal force exerted on 1 is $ F_1=- \partial_{d_1}\mathcal E$ where $\mathcal E$ is the Casimir interaction energy. Following the line of logic of the previous section, this force can be very well described by the lowest order diagrams. However here we are interested in how the force on 1 changes upon changing the position of 2 which we can define by $I_{12}\equiv\partial_{d_2}F_1=-\partial_{d_1}\partial_{d_2} \mathcal E$.
\begin{figure}[h]
\centering
\def\svgwidth{6cm}
\subfigure[ { Cancelation of the two-body and three-body diagrams.\label{Fig-CancellationofTwoAndThreeBodyDiagram} }]{
 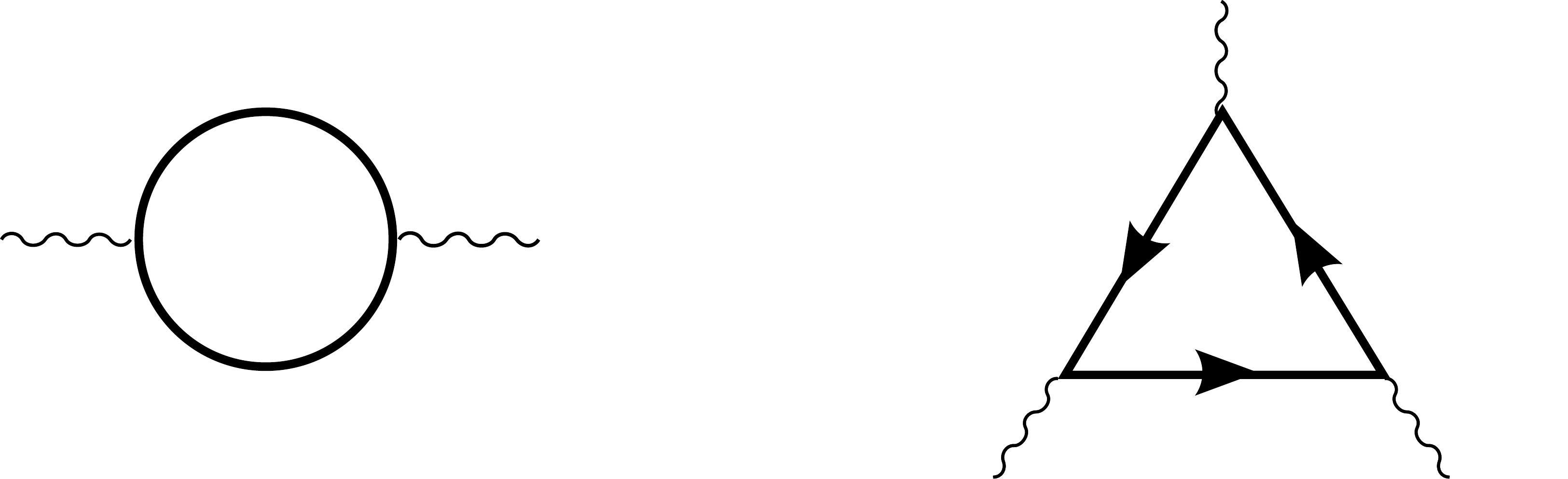
}\qquad
\def\svgwidth{6.5cm}
\subfigure[{{ Cancelation of the (other) three-body diagram against a fourth-order diagram.\label{Fig-CancellationofFourAndThreeBodyDiagram} }}]{
 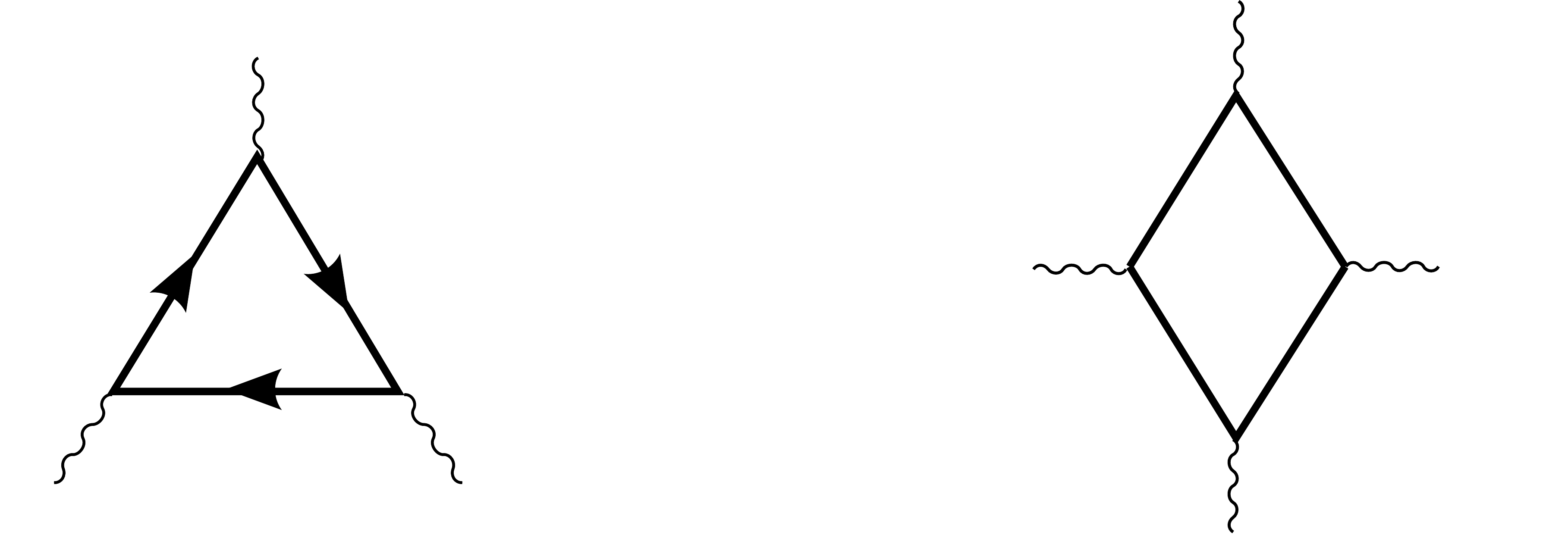
}
\caption{\label{Fig-Cancellation}{In the extreme limit that the object 3 is an infinite plate, the interaction diagrams cancel each other in pairs. If the blocking is partial, the interaction between the objects 1 and 2 is partially screened.}}
\end{figure}
Note that in the extreme case that the vertical half-plate completely blocks the space between 1 and 2, $I_{12}$ just vanishes. Depending on whether the blocking is partial or complete, the interaction is partially or completely screened. We will see that multiple-reflection expansion provides a natural language to better understand this.

Let us first consider the extreme case where the vertical half-plate is almost completely blocking the way from 1 to 2. The half-plate can be then regarded as an infinite plate.
Indeed, in this limit the total energy is just the sum of the interaction of the infinite plate with either of the two other objects. In the diagrammatic expansion, however, there is a direct two-body interaction between 1 and 2. But one can see that this diagram is canceled against the three-body diagram which connects the three objects (see
Fig.~\ref{Fig-CancellationofTwoAndThreeBodyDiagram}): The mathematical expression for these diagrams are given by
\begin{align}
 & \mfmadd{[21]}
 =-\frac{\hbar c}{2\pi}\int_{0}^{\infty} d\kappa \, {\rm tr} \left(\mathcal{U}_{12}T_2\, \mathcal{U}_{21}\,T_1\right), \\
 & \mfmadd{[321]}
 =-\frac{\hbar c}{2\pi}\int_{0}^{\infty} d\kappa \, {\rm tr} \left(T_1 \, \mathcal{U}_{12}T_2\, \mathcal{U}_{23} T_3 \, \mathcal{U}_{31} \label{Eq. three body diagram}\right).
\end{align}
In the last equation we are concerned with the RL channel of $T_3$ (the $T$-matrix of the infinite plate). For an infinite plate, this is negative unity matrix  for both Dirichlet and Neumann boundary conditions
\begin{equation}
  T^{\substack{\scriptsize \rm Infnite\\ \rm -plate} }_{RL }=-\mathcal I.
\end{equation}
By implementing this in Eq.~(\ref{Eq. three body diagram}) and noting that $\mathcal{U}_{13}\, \mathcal{U}_{32}=\mathcal{U}_{12}$, we find that \mfmadd{ $[21]+[321]=0$}.
The cancelation between these types of diagrams can be carried out to all orders of the diagrammatic expansion (see Fig.~\ref{Fig-Cancellation}).
\begin{figure}[h]
  \centering
  \def\svgwidth{7cm}
  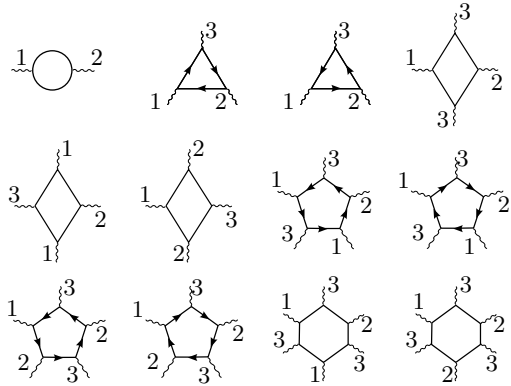
\caption{\label{Fig-DiagramsContributeToBlocking} The lowest order diagrams in reflections contributing to the interaction between the horizontal half-plates. \mfmadd{The diagrams are sorted by the typical optical paths that the waves travel between the objects. The interaction is dominated by the diagrams in the first row. The rest of the diagrams only slightly modify the leading order contribution}.}
\end{figure}

\begin{figure}[b]
  \def\svgwidth{10cm}
  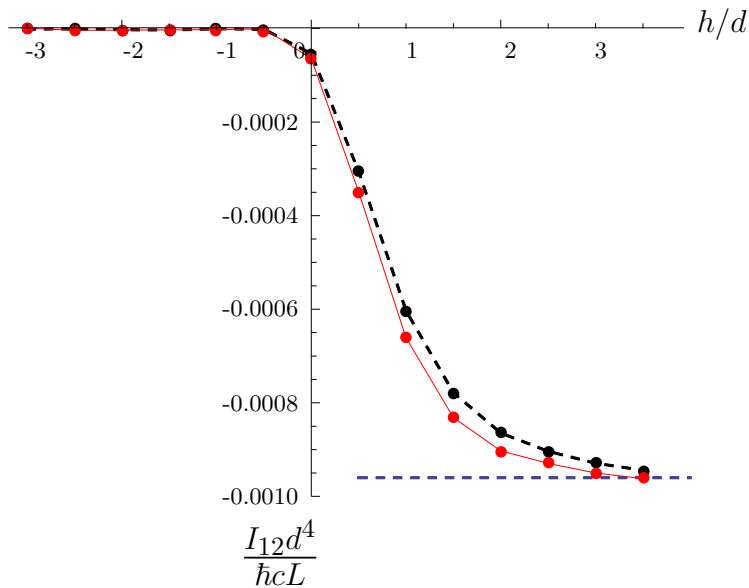
\caption{The interaction of the two horizontal half-plates as a function of the position of the blocking half-plate. \mfmadd{The dashed curve is the resultant force by considering the first few diagrams in the reflection expansion. The solid curve includes the correction of higher orders in reflections. The blue dashed line represents the constant to which the solid curve approaches as $h\to\infty$.} \label{Plot-Blocking} }
\end{figure}

When the vertical half-plate is far above the other two, the higher order diagrams do not contribute and thus there is no screening. Therefore, it is conceivable that the sum of the diagrams in Fig.~\ref{Fig-Cancellation} gives a first approximation to $I_{12}$. We will see that this intuition is qualitatively correct but higher orders must be taken into account to improve the quantitative result.

We compute the interaction $I_{12}$ by including all the diagrams shown in {Fig.~\ref{Fig-DiagramsContributeToBlocking}}.
This quantity is plotted in  Fig.~\ref{Plot-Blocking} as the solid (red) curve. The dashed (black) curve is the resultant interaction when only the diagrams in Fig.~\ref{Fig-Cancellation} are included.
Again we see that the multiple reflections enjoy a rapid convergence.

The interaction $I_{12}$ monotonically decreases as the vertical half-plate further blocks the two objects. Interestingly, as it approaches them, the interaction falls off very rapidly. Even before the vertical half-plate reaches the axis along the horizontal half-plates, the interaction is substantially suppressed. {After reaching this point, $I_{12}$ falls off sharply to zero. This suggests that the interaction of two objects is greatly affected by whether or not they are visible to each other}.

\subsection{Interaction of edges and a needle}\label{Sec. Interaction of edges and dipoles}
\begin{figure}[h]
  \centering
  \def\svgwidth{12cm}
  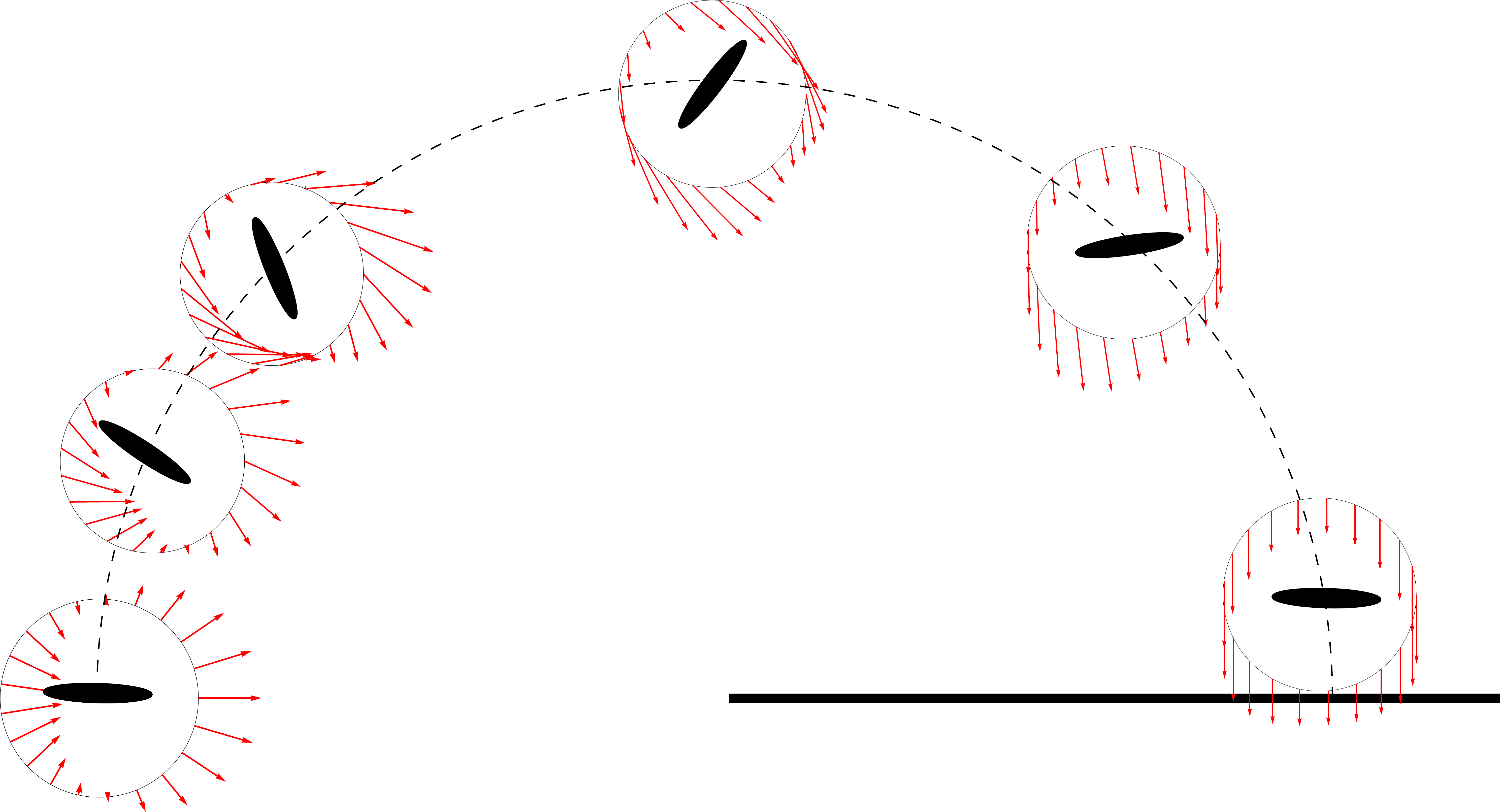
\caption{The direction (and the normalized magnitude) of the force for $\phi_0=0, \pi/8, \pi/4, \pi/2,3\pi/4$ and $\pi^-$. {See text for further description.}  \label{Fig-ForceYYDirection} }
\end{figure}

{In this section, we study the interaction between edges and a needle. In Sec.~\ref{Sec. Edge vs dipole}, we consider an edge and a needle in two dimensions and find {that their interaction} exhibits some unusual features.
Using these results, in Sec.~\ref{Sec. Repulsion}, we will find analytical results for an example of repulsion which was first proposed in Ref.~\cite{Levin10}.

\subsubsection{Edge vs needle}\label{Sec. Edge vs dipole}
In a recent paper \cite{Levin10}, a setup was proposed which gives rise to a repulsive Casimir force. \mfmadd{This setup consists of two objects which can be separated by an imaginary plane.} A two dimensional model of the repulsion \mfmadd{involves }a small elliptical perfect metal above a (perfect) metal line with a gap (see {Fig.~\ref{Fig-EllipseTwoHalfLines}}). The Casimir force is then computed based on 2d electromagnetism.

\mfmadd{As formulated in the Appendix, the electromagnetism in two dimensions (two spatial dimensions plus time) }with perfectly-conducting boundary condition is equivalent to a scalar field theory with Neumann boundary condition. We exploit this fact to compute the interaction between an edge and a needle (Fig.~\ref{Fig-EllispeHalfLine}).

The interaction can be expanded in multiple reflections as in Fig.~\ref{Fig-TwoBodyDiagExp}. The lowest order will suffice because the ellipse (needle) is small compared to the separation distance \cite{Casimir48-1}.
In order to compute this diagram, we must know scattering properties of these objects. The half-plate's $T$-matrix is given by Eq.~(\ref{Eq. half plate T-matrix}). The $T$-matrix of the ellipse can be described in the cylindrical coordinates where a natural basis is $H^{(1)}_m(kr)e^{i m\phi}$. The scattering matrix is then labeled in this basis, hence $T_{m,m'}$. In the limit of a small ellipse with Neumann boundary condition we need a subset $m,m'\in\{0,\pm 1\}$ of the scattering matrix. This is further simplified by noting the inversion symmetry of the ellipse which requires $m+m'$ to be even; so $m,m'=0$ is decoupled from $m,m'=\pm 1$. Therefore, the only nonzero components of the $T$-matrix in this subset are $T_{00}$ and $T_{m,m'}$ for $m,m'=\pm1$. Using parity, the latter components of the $T$-matrix (with $m,m'= \pm 1$) can be written as a superposition of $T_{xx}$ and $T_{yy}$ where $x$ and $y$ are the symmetry axes of the ellipse. For a small object, we need the $T$-matrix to be expanded to the lowest order in the wave number $\kappa$, in which limit they all depend quadratically on $\kappa$, that is $T_{00}\sim\kappa^2 \mathcal T_{00},T_{xx}\sim\kappa^2 \mathcal T_{xx},T_{yy}\sim\kappa^2 \mathcal T_{yy}$. We leave the $\mathcal T$'s as parameters without {specializing to} an ellipse ---the discussion applies to any geometry with the same symmetry properties as an ellipse.

The $T$-matrix of the half-line is given in planar basis while that of the needle is defined in the cylindrical basis. We can convert the two bases by
\begin{equation}
  e^{i\mathbf k.\mathbf x}=e^{-\kappa r \cos(\phi-a)}=\sum_{m=-\infty}^{\infty} (-1)^m I_m(\kappa r) e^{im(\phi-a)},
\end{equation}
where $\phi$ and $a$ are the angles of $\mathbf x$ and $\mathbf k$ respectively. So the {\it conversion} matrix is $D_{a, m}=(-1)^m e^{-i m a}$. The $T$-matrix of the needle can be then cast in the planar basis using the conversion matrix and the normalization factors which are involved in the definition of these functions \cite{Rahi09}
\begin{equation}
  T_{a',a}= \pi \sum_{m,m'} (-1)^{m+m'}e^{i m' {a'}^*-ima } T_{m,m'}.
\end{equation}

Now we can put everything together to compute the Casimir interaction energy of a half-line and a needle. \begin{figure}[h]
  \centering
  \def\svgwidth{8cm}
  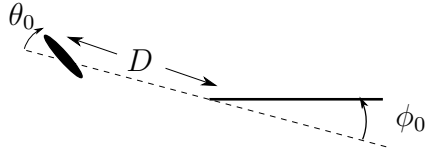
\caption{The configuration a needle versus a half-line.\label{Fig-EllispeHalfLine} }
\end{figure}
The objects are separated by a distance $D$ along the needle-edge axis with respect to which the needle makes an angle $\theta_0$ and the half-line is tilted at an angle $\phi_0$ (see Fig.~\ref{Fig-EllispeHalfLine}). The energy finds contributions from different components of the $T$-matrix. For each contribution to the Casimir interaction, we can find exact analytical formulae\footnote{Higher orders are suppressed by higher powers of the needle dimension divided by the separation distance.}. The term proportional to $\mathcal T_{00}$ is
\begin{equation}
    \mathcal E_{00}\sim -\frac{\hbar c}{64\pi}\frac{1}{D^3} \mathcal T_{00} \left(-\frac{4}{3}+\csc^3\phi_0 \left(2\phi_0-\sin (2\phi_0)\right)\right).
\end{equation}
\mfmadd{where $\sim$ indicates that this result is exact in the limit of a vanishing needle.}
It is not surprising that this expression does not depend on $\theta_0$ since the lowest partial wave (which is denoted by 0) does not detect the orientation of the needle. The dependence on the separation distance is obvious for dimensional reasons. The dependence on the tilt angle, however, is nontrivial. Interestingly when the half-line is aligned with the axis which connects it to the needle, this term becomes zero---irrespective of the separation distance. {We will discuss this in more detail in the next section.}

The dependence on the orientation of the needle comes through $\mathcal T_{xx}$ and $\mathcal T_{yy}$ and is given by
\begin{align}\label{Eq. Exx}
  \mathcal E_{xx}\equiv \mathcal T_{xx} \, f(\phi_0, \theta_0)\nonumber \\
  \sim  -\frac{\hbar c}{8\pi}\frac{1}{D^3} \mathcal T_{xx} \Big(&-\frac{4}{3}+\cos(2\theta_0)\left(-2+\cot\phi_0\csc\phi_0\right)\nonumber\\
  &-\csc \phi_0\left(3\cot\phi_0+2\sin(2\theta_0)+\phi_0(-3+\cos(2\theta_0+2\phi_0))\csc^2\phi_0)\right)\Big),
\end{align}
and, \mfmadd{by symmetry,}
\begin{align}\label{Eq. Eyy}
  \mathcal E_{yy} (\phi_0, \theta_0)=   \mathcal T_{yy} \, f\left(\phi_0, \theta_0+\frac{\pi}{2}\right).
\end{align}
The sum of these expressions give the interaction of a needle with a half-line. The last equation is a consequence of the symmetries of the problem. Note that these expressions all come with the same dependence on the separation distance; so in principle they are all important at the leading order. A simple consistency check is that the sum $f\left(\phi_0, \theta_0\right)+f\left(\phi_0, \theta_0+\frac{\pi}{2}\right)$ must be independent of the orientation $\theta_0$. That is because it gives the interaction of a circle---as opposed to an ellipse---with the half-line.

\begin{figure}[h]
\centering
 \def\svgwidth{5.5cm}
\subfigure[ { The magnitude of the force for $\phi_0=0$. The force vanishes \mfmadd{(at least, in the lowest reflection)} for $\theta_0=\pi/2$ at all separation distances.  \label{Plot-ForceYY} }]{
  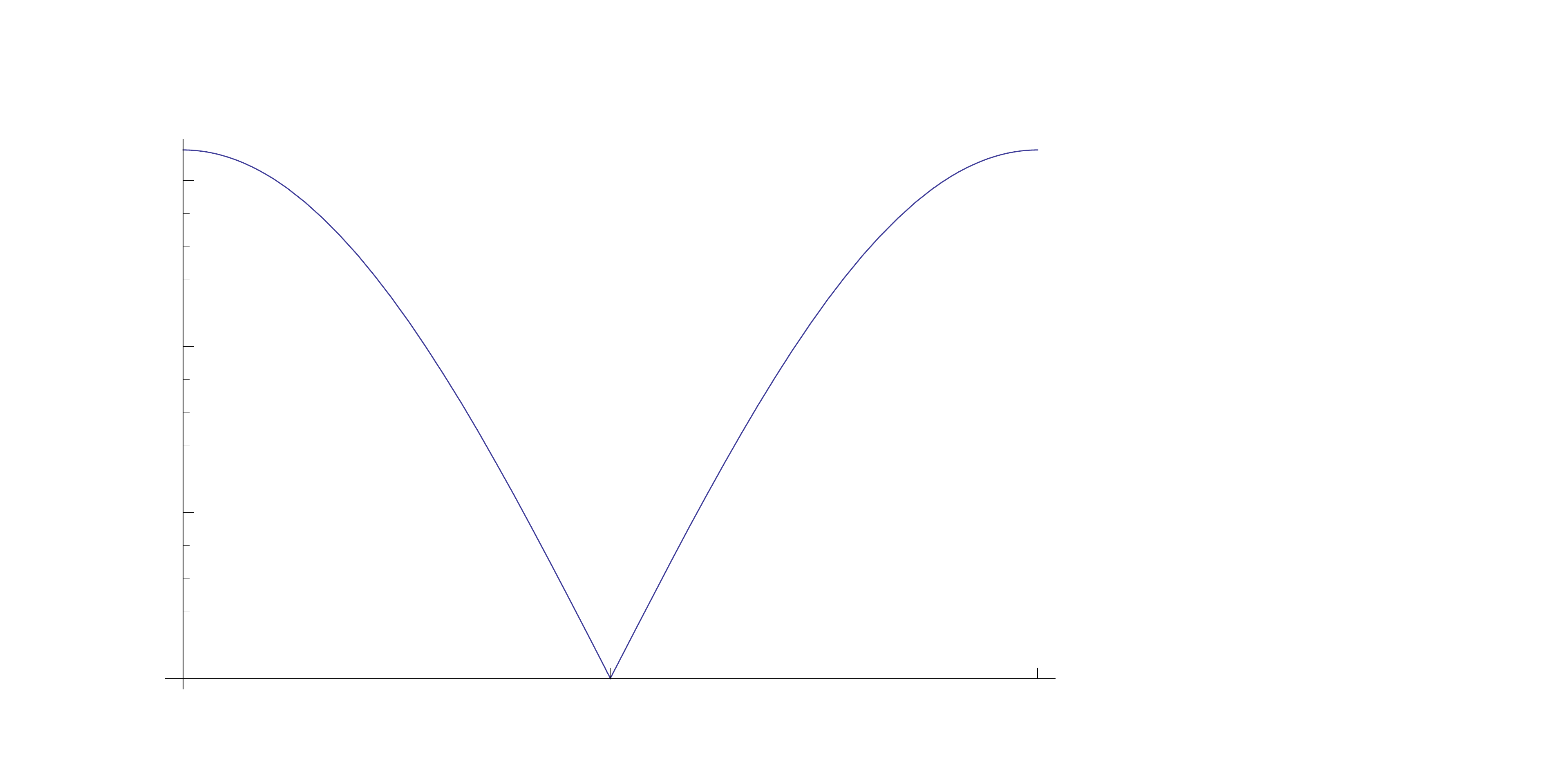
}\qquad
 \def\svgwidth{8.3cm}
\subfigure[ { The magnitude of the force for different orientations and positions of the needle including $\phi_0=0$. From bottom to top, the curves represent $\phi_0=0, \pi/8, \pi/4$ and $\pi/2$ }.\label{Plot-ForceYY2}]{
  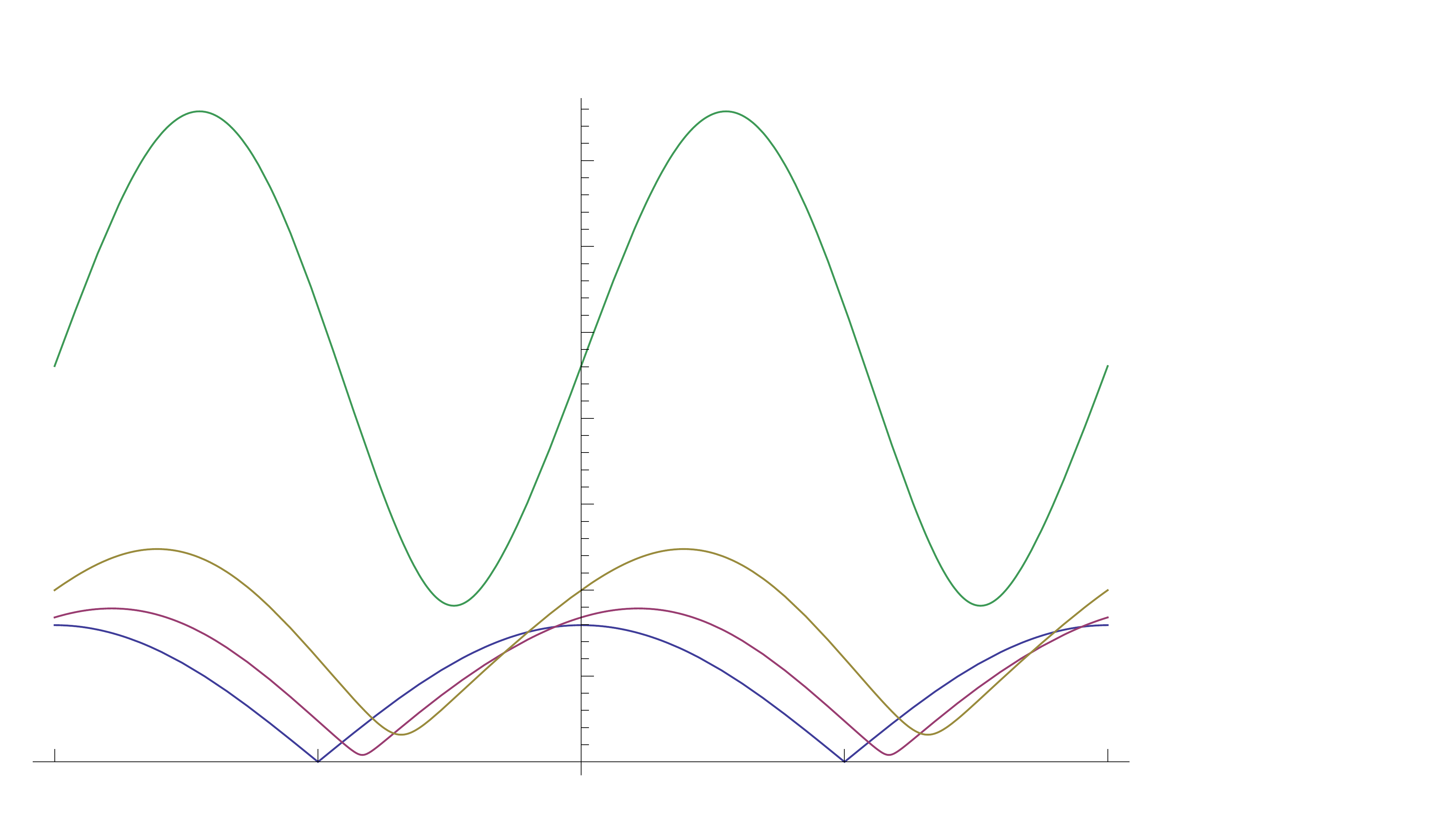
}
\label{?}
\caption{{The magnitude of the force as a function of the needle orientation. The force strongly depends on the orientation.}}
\end{figure}

It is interesting that the complicated interaction of a {\it needle} with an {\it edge} can be described analytically. The nontrivial dependence on the mutual orientation can be studied at different values and limits of the angles.
In the limit that $\phi_0$ approaches $\pi$, the interaction becomes that of an infinite line with a {\it dipole}. A closely related problem of a dipole opposite an infinite conducting plate has been studied in Ref.~\cite{Emig09} {in three dimensions}. In {another} limit, $\phi_0=0$, the needle is aligned along the axis of the half-line{,} in which case the orientation-dependent terms, $\mathcal E_{xx}$ and $\mathcal E_{yy}$, exhibit unusual characteristics as a function of the orientation \mfmadd{which will be discussed in more detail below}. Note that in the same limit, $\mathcal E_{00}$ vanishes.

Here we assume that the needle is elongated in one direction. Choosing this axis to be $y$, we can then neglect $\mathcal E_{xx}$. Also the contribution from $\mathcal E_{00}$ is significantly smaller than the orientation dependent terms; so the total energy is given by $\mathcal E_{yy}$ without any loss of generality. The absolute value of the force (derived from Eqs.~(\ref{Eq. Exx}) and (\ref{Eq. Eyy})) as a function of the needle orientation for $\phi_0=0$ is plotted in Fig.~\ref{Plot-ForceYY}.
The force is symmetric with respect to $\pi/2$ as it should be. But it also vanishes at this point. This was first observed in Ref.~\cite{Levin10} where it is explained by employing an intuitive argument.
Figure~\ref{Plot-ForceYY2} gives the magnitude of the force for a set of different angles. Away from $\phi_0=0$ the force is non-vanishing but it shows a strong dependence on the angle $\theta_0$.

The direction of the force also exhibits some interesting features. Let us consider several points all at the same distance from the edge. We } show the direction (and the {relative} magnitude of the force) in a small circle around each point (see Fig.~\ref{Fig-ForceYYDirection}).
An arrow drawn from a point on the circle should be understood as the force on the needle which is pointed from the center of the circle to that point.
The force is normalized within each circle but differently from other circles. The small needle drawn in each circle is the equilibrium configuration at each point.

This plot illustrates some characteristics of the interaction of an {\it edge} with a {\it needle}:

1. There is a strong dependence on the orientation of the needle. Both the magnitude and direction of the force vary strongly. In fact, when $\phi_0=0$ the {\it dipole} drags the force direction along with itself (parallel to its orientation).

2. The vertical component of the force does not always point towards the half-line. For $\phi_0=\pi/8$, for example, we can see that for a certain range of the orientation $\theta_0$, the (vertical) force points upward. {This will be essential to the argument in the next section.}

3. The equilibrium configuration (the point of zero torque) strongly depends on $\phi_0$, the relative positioning of the two objects.

\subsubsection{Repulsion}\label{Sec. Repulsion}
\begin{figure}[ht]
  \centering
  \def\svgwidth{9cm}
  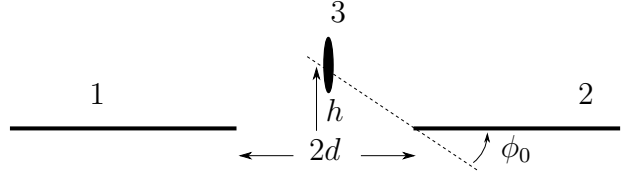
\caption{\label{Fig-EllipseTwoHalfLines} {The configuration of a needle against a {\it gap}}. }
\end{figure}

In this section, we follow the setup in Ref.~\cite{Levin10} and
consider the interaction between a vertical needle and a line with a gap, {\it or} two half-lines (see Fig.~\ref{Fig-EllipseTwoHalfLines}).

\begin{figure}[h]
  \centering
  \def\svgwidth{8cm}
  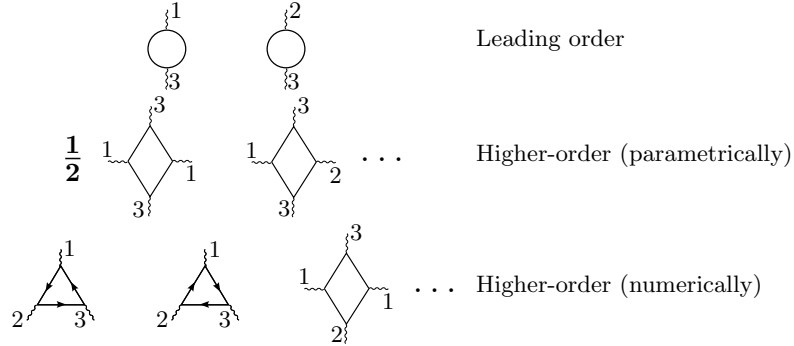
\caption{\label{Fig-RepulsionDiagrams} {Diagrammatic expansion in multiple reflections. The two-body diagrams dominate the interaction. The higher-order diagrams with more than one needle vertex are parametrically smaller. Other higher-order diagrams are numerically smaller due to the rapid convergence in higher reflections.}}
\end{figure}

\begin{figure}[b]
  \centering
  \def\svgwidth{9cm}
  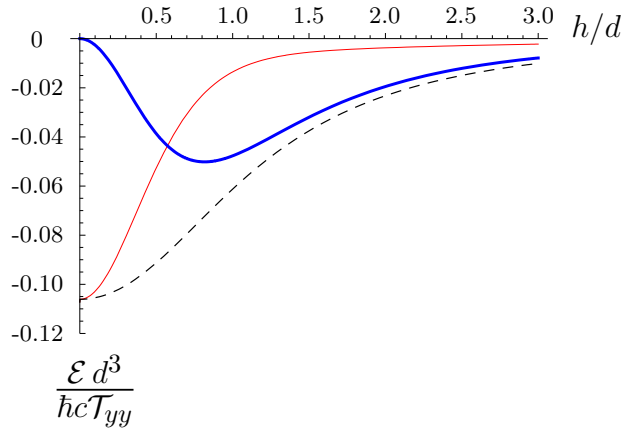
\caption{{Casimir interaction energy of the vertical and horizontal needle and the circle with the gap in the leading order of multiple reflections. \mfmadd{The bold (blue) curve represents the energy of a vertical needle, the fine (red) curve gives the energy of a horizontal needle and the dashed curve is the energy of a circle opposite the gap.}} \label{Plot-RepulsionEnergy} }
\end{figure}

The two-body interaction of a half-line with an ellipse was computed in the previous section. Here we also consider higher refections. Because the ellipse is small in size, it will suffice to consider only a single reflection off of it. But higher reflections {from} the half-lines should be included.
The three-body diagram in {Fig.~\ref{Fig-RepulsionDiagrams}} is an example.
We will find that the latter diagrams are also significantly smaller than the two-body diagrams. This again proves the {usefulness} of the expansion in multiple reflections.
Figure~\ref{Fig-RepulsionDiagrams} summarizes these statements.
As a result, we can very precisely describe the interaction as the sum of the two-body diagrams. Using Eqs.~(\ref{Eq. Exx}) and (\ref{Eq. Eyy}), the energy takes the form \begin{equation}
  \mathcal E \sim \mathcal T_{yy}\, \frac{\cot^3\phi_0  }{48 \pi d^3 } \, \Big(-24 \phi_0+6 \sin(2 \phi_0 )+5 \sin(3 \phi_0 ) +3\sin(4 \phi_0 )-3\sin(5 \phi_0 )\Big).
\end{equation}
This expression is plotted as the bold (blue) curve in Fig.~\ref{Plot-RepulsionEnergy} where it has been also
compared with the case where the needle was positioned horizontally ({red curve}). The {dashed curve} belongs to a circle of the same diameter as the needle and is simply obtained by summing the values of the other two curves. \mfmadd{Note that the contribution due to $\mathcal T_{00}$ is smaller by one order magnitude which is why it is neglected.}
\begin{figure}[h]
  \centering
  \def\svgwidth{9cm}
  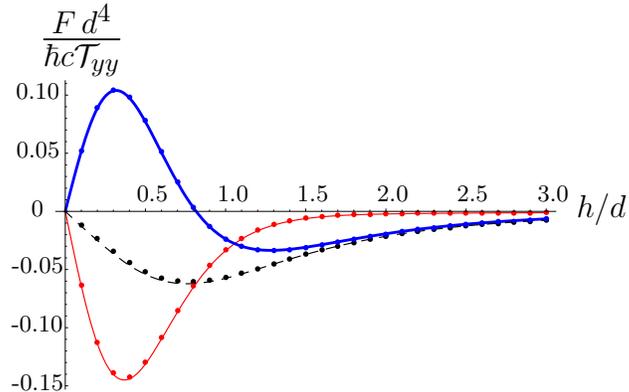
\caption{{Casimir force on the vertical and horizontal needle and the circle with the gap. They are plotted as the bold (blue), fine (red) and the dashed curves respectively. The dots include the correction from higher reflections.}}\label{Plot-RepulsionForce}
\end{figure}
The interesting point about the {bold curve}, also discussed in the previous section, is that it vanishes for $h=0$. The energy, however, is always negative so the force on a vertical needle in the proximity of the gap is actually repulsive. This observation has been used to propose a mechanism of obtaining repulsion in Ref.~\cite{Levin10}. On the other hand, the force on a horizontally-positioned needle or a circle is never repulsive. These are plotted in Fig.~\ref{Plot-RepulsionForce}.
\mfmadd{We also include the corrections to the force due to higher order reflections. }The dots include the first correction from the three-body diagrams which is extremely small compared to the leading order (shown by the solid and dashed curves).  \mfmadd{Again this shows that a systematic expansion of multiple reflections gives a remarkably accurate result.}

\mfmadd{The configuration in Fig.~\ref{Fig-EllipseTwoHalfLines} gives rise to repulsion, but it is unstable with respect to the rotations in the plane. Also according to the theorem in Ref.~\cite{Rahi10}, it should be unstable with respect to the displacement in the horizontal direction.}

\mfmadd{
\section{Conclusions}
In this work, we have developed a multiple-reflection expansion of the Casimir interaction. We have represented the latter in a diagrammatic fashion and derived simple rules to organize and consistently expand it.

We have argued that the formal diagrammatic expansion converges rapidly and thus it is sufficient to keep only the lowest order diagrams in the number of reflections. We have explicitly shown the latter for various configurations. Specifically, the interaction between edges (of half-plates) among themselves and with a needle has been computed. Most notably, we have found an analytical expression describing the interaction of two half-plates in the lowest reflection, that should be accurate for practical applications. Three-body interactions of half-plates are considered where we have taken advantage of the remarkable convergence in multiple reflections. Finally the interaction between edges and a needle is studied to find analytical formulae. The results have been used for an analytic treatment of a model of repulsion which has been proposed recently.

A more rigorous bound on the convergence of higher order diagrams is still desired for both two-body and multibody configurations. The expansion in the lowest reflections can be a basis for both numerical and analytical studies of complex multibody configurations and geometries. The rapid convergence in diagrammatic expansion can lead to a remarkable simplification of the intensive computations.

}
\begin{acknowledgements}
  This research was supported by the NSF Grant No.
DMR-08-03315, DARPA contract No. S-000354, and the U. S. Department of
Energy (DOE) under cooperative research agreement \#DF-FC02-94ER40818. I wish to thank Robert L. Jaffe, Mehran Kardar and Matthias Kr{\"{u}}ger for many valuable discussions. I am specially grateful to Prof. Jaffe and Prof. Kardar for a critical reading of the manuscript.
\end{acknowledgements}

\section*{Appendix: Electromagnetism in two dimensions}

In any dimensions, the Maxwell equations can be written in their most general form as a tensor equation
\begin{equation}
  \partial_\mu F^{\mu \nu} =0,
\end{equation}
where $F^{\mu\nu}$ is the field strength tensor and $\mu,\nu=0,1,2,...$ indicate the space-time coordinates.
In two dimensions, this reads
\begin{align}
 &\partial_1 F^{10}+\partial_2 F^{20}=0,\nonumber \\
 &\partial_0 F^{01}+\partial_2 F^{21}=0,\nonumber\\
 &\partial_0 F^{02}+\partial_1 F^{12}=0.
\end{align}
We can solve these coupled equations by introducing an auxiliary field $\psi$ such that
\begin{align}
  &F^{10}=\partial_2\psi, \nonumber\\
  &F^{20}=-\partial_1 \psi, \nonumber\\
  &F^{12}=-\partial_0 \psi.
\end{align}
The {\it gauge} freedom has been fixed by this set of choices. The electromagnetic lagrangian $-\frac{1}{4} F_{\mu\nu}F^{\mu\nu}$ then becomes proportional to
\begin{equation}
  \frac{1}{2}\left((\partial_0 \psi)^2-(\nabla\psi)^2\right)\,,
\end{equation}
which is the lagrangian of a scalar field.

The boundary condition for a perfect conductor can be written in a covariant form too
\begin{equation}
  \epsilon_{\mu \nu \rho} F^{\mu\nu}n^{\rho}{\large|}_{\Sigma}= 0\,,
\end{equation}
where $\Sigma$ denotes the surface of the object. In this equation, $\epsilon$ is the completely antisymmetric tensor and $n$ is the vector normal to the (hyper-)surface of the object. When the auxiliary field is replaced for the field tensor, the last equation becomes
\begin{equation}
  \partial_n \psi=0.
\end{equation}
Hence, the quantum electrodynamics in two spatial dimensions with perfect conductors as the boundaries is equivalent to a quantum scalar field theory with Neumann boundary condition.

\end{document}